\documentclass[]{aastex}

\usepackage{emulateapj5}
\usepackage{onecolfloat}
\usepackage{graphicx} 
\usepackage{fancyheadings} 
\usepackage{ulem}
\usepackage{rotating}
\usepackage{lscape}

\def\lsim{\hbox{ \rlap{\raise 0.425ex\hbox{$<$}}\lower 0.65ex\hbox{$\sim$}}}
\def\gsim{\hbox{ \rlap{\raise 0.425ex\hbox{$>$}}\lower 0.65ex\hbox{$\sim$}}}
\def\ale{\mathrel{\hbox{\rlap{\hbox{\lower4pt\hbox{$\sim$}}}\hbox{$<$}}}}
\def\age{\mathrel{\hbox{\rlap{\hbox{\lower4pt\hbox{$\sim$}}}\hbox{$>$}}}}

\newcommand{\msol}{M_\odot}
\newcommand{\kms}{\mbox{km\ s${^{-1}}$}}
\def\arcmin{\hbox{$^\prime$}}
\def\arcsec{\hbox{$^{\prime\prime}$}}
\def\fd{\hbox{$~\!\!^{\rm d}$}}
\def\fh{\hbox{$~\!\!^{\rm h}$}}
\def\fm{\hbox{$~\!\!^{\rm m}$}}
\def\fs{\hbox{$~\!\!^{\rm s}$}}
\def\keck{1}
\def\lick{2}
\def\berk{3}
\def\uc{4}
\def\ias{5}
\def\chandra{6}
\def\kipac{7}
\def\mex{8}
\def\hubb{9}
\def\ssl{10}
\def\harvard{11}
\def\phys{12}
\def\ucla{13}
\def\lbl{14}
\def\ubc{15}
\def\ucd{16}
\def\llnl{17}
\def\jpl{18}

\begin{document}

\twocolumn[%
\submitted{Accepted to ApJ: January 2006}

\title{The Galaxy Hosts and Large-Scale Environments of 
Short-Hard $\gamma$-ray Bursts}

\author{J. X. Prochaska\altaffilmark{\keck,\lick}, 
        J. S. Bloom\altaffilmark{\berk},
        H.-W. Chen\altaffilmark{\uc}, 
	R. J. Foley\altaffilmark{\berk}, 
        D. A. Perley\altaffilmark{\berk},
	E. Ramirez-Ruiz\altaffilmark{\ias,\lick,\chandra}, 
        J. Granot\altaffilmark{\kipac,\ias},
        W. H. Lee\altaffilmark{\mex},
        D. Pooley\altaffilmark{\berk,\hubb},
        K. Alatalo\altaffilmark{\berk},
        K. Hurley\altaffilmark{\ssl},
        M. C. Cooper\altaffilmark{\berk}, 
        A. K. Dupree\altaffilmark{\harvard}, 
        B. F. Gerke\altaffilmark{\phys},
        B. M. S. Hansen\altaffilmark{\ucla},
        J. S. Kalirai\altaffilmark{\lick},
        J. A. Newman\altaffilmark{\lbl,\hubb},
        R. M. Rich\altaffilmark{\ucla},
        H. Richer\altaffilmark{\ubc},
	    S. A. Stanford\altaffilmark{\ucd,\llnl},
	    D. Stern\altaffilmark{\jpl},
	    W.J.M. van Breugel\altaffilmark{\llnl}
	}

\begin{abstract} 
The rapid succession of discovery of short--duration hard--spectrum GRBs 
has led to unprecedented insights into the energetics of the explosion 
and nature of the progenitors. Yet short of the detection of a 
smoking gun, like a burst of coincident gravitational radiation 
or a Li-Paczy\'{n}ski {\em mini-supernova}, it is unlikely 
that a definitive claim can be made for the progenitors. As was the 
case with long--duration soft--spectrum GRBs, however, the expectation is 
that a systematic study of the hosts and the locations of short GRBs 
could begin to yield fundamental clues about their nature.  We 
present an aggregate study of the host galaxies of short--duration 
hard--spectrum GRBs. In particular, we present the Gemini--North and Keck 
discovery spectra of the galaxies that hosted three short GRBs and a 
moderate--resolution ($R\approx 6000$) spectrum of a fourth host. We 
find that these short--hard GRBs originate in a variety of low-redshift
($z<1$) environments that 
differ substantially from those of long--soft GRBs, both on individual 
galaxy scales and on galaxy--cluster scales. Specifically, three of the 
bursts are found to be associated with old and massive galaxies with 
no current ( $< 0.1 \msol$\,yr$^{-1}$) or recent star formation. 
Two of these galaxies are located within a cluster environment. These 
observations support an origin from the merger of compact stellar remnants, 
such as double neutron stars of a neutron star--black hole binary. The 
fourth event, in contrast, occurred within a dwarf galaxy with a star 
formation rate exceeding 0.3~$\msol$\,yr$^{-1}$. Therefore, it 
appears that like supernovae of Type Ia, the progenitors of short--hard 
bursts are created in all galaxy types, suggesting a corresponding 
class with a wide distribution of delay times between formation and 
explosion. 


\end{abstract}
]

 \altaffiltext{\keck}{Visiting Astronomer, W. M. Keck Telescope.
 The Keck Observatory is a joint facility of the University
 of California and the California Institute of Technology.}
 \altaffiltext{\lick}{UCO/Lick Observatory, University of California, Santa Cruz; 
   Santa Cruz, CA 95064}
 \altaffiltext{\berk}{Department of Astronomy, 601 Campbell Hall, 
        University of California, Berkeley, CA 94720-3411}
 \altaffiltext{\uc}{Department of Astronomy and Astrophysics, University of Chicago, Chicago, IL 60637}
 \altaffiltext{\ias}{Institute for Advanced Study, Olden Lane, Princeton, NJ 08540}
 \altaffiltext{\chandra}{Chandra Fellow}
 \altaffiltext{\kipac}{KIPAC, Stanford University, P.O. Box 20450, Mail Stop
    29, Stanford, CA 94309}
 \altaffiltext{\mex}{Instituto de Astronomia, UNAM Apdo. Postal 70-264 Mexico DF 04510 Mexico}
 \altaffiltext{\hubb}{Hubble Fellow}
 \altaffiltext{\ssl}{UC Berkeley, Space Sciences Laboratory, 7 Gauss Way, 
	Berkeley, CA 94720-7450}
 \altaffiltext{\harvard}{Harvard-Smithsonian Center for Astrophysics, 60 Garden St., 
	Cambridge, MA 02138}
 \altaffiltext{\phys}{Department of Physics, 366 LeConte Hall, University of
      California, Berkeley, CA 94720-7300}
 \altaffiltext{\ucla}{Department of Physics and Astronomy, 
	University of California Los Angeles, Los Angeles, CA, 90095}
 \altaffiltext{\lbl}{Institute for Nuclear and Particle Astrophysics,
  	Lawrence Berkeley National Laboratory, Berkeley, CA 94720}
 \altaffiltext{\ubc}{Physics \& Astronomy Department, University of British Columbia,
	Vancouver, B.C. V6T 1Z1}
 \altaffiltext{\ucd}{Department of Physics, One Shields Ave., University
        of California, Davis, CA 95616-8677}
 \altaffiltext{\llnl}{Institute for Geophysics and Planetary Physics,
	Lawrence Livermore National Laboratory, L-413
	7000 East Ave, Livermore CA 94550}
 \altaffiltext{\jpl}{Jet Propulsion Laboratory, California Institute of Technology, 
	4800 Oak Grove Drive, MS 169-506, Pasadena, CA 91109}

\pagestyle{fancyplain}
\lhead[\fancyplain{}{\thepage}]{\fancyplain{}{PROCHASKA ET AL.}}
\rhead[\fancyplain{}{Galaxy Hosts and Large-Scale Environments of 
Short-Hard $\gamma$-ray Bursts}]{\fancyplain{}{\thepage}}
\setlength{\headrulewidth=0pt}
\cfoot{}

\section{Introduction}

The nature of the progenitors of short-duration, hard
spectrum, gamma-ray bursts \citep{kmf+93} (GRBs) has remained a mystery. 
Even with the recent localizations of four short-hard
GRBs, no transient emission has been found at long wavelengths 
that directly constrains the progenitor nature. 
Instead, as was the case in studying the different morphological subclasses of
supernovae \citep{rea53,vandyk92} and the progenitors of long-duration
GRBs \citep{bkd02}, here we argue that the progenitors of short bursts can
be meaningfully constrained by the environment in which the bursts occur.  

\begin{figure*}
\begin{center}
\includegraphics[angle=-90,width=6.3in]{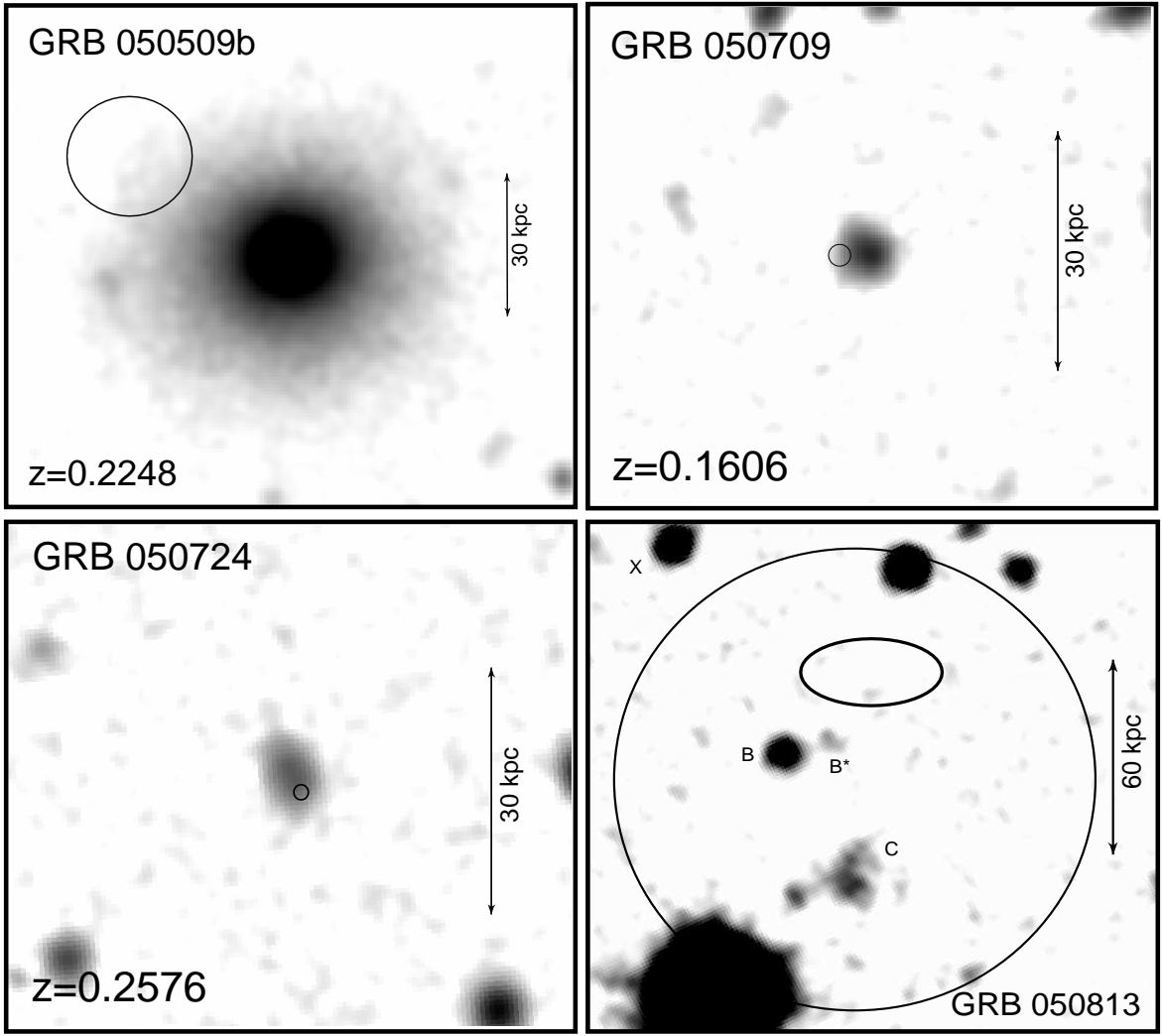}
\caption{Optical light montage of four host galaxy regions of 
short-hard GRBs.  
The ellipses in each panel represent the astrometric position of the 
most accurate X-ray afterglow position reported with the
exception of GRB~050813 (see text).
In the case of GRB~050709 and GRB~050724 where 
optical afterglows were detected, the GRB is projected to within 
2\arcsec\ from the center of a galaxy with apparent magnitude 
$R< 19.5$\,mag. The likelihood of a chance association between 
these afterglows and the putative host galaxies is less than 
10$^{-4}$ per event given the covering fraction of such objects 
on the sky. Similarly, the error circle containing GRB~050509b 
encompasses a single bright
galaxy which is the putative host galaxy \citep{bpp+05} for which the
chance of a spurious physical association with the burst is $\sim
10^{-3}$. 
Adopting the redshift of the putative host or cluster 
redshift (GRB\,050813) a projection scale is shown at right in each panel.  
All images were smoothed with a Gaussian of 1.4--1.6 pixels to enhance 
the contrast between detected objects and sky noise. North is up and 
East is to the left.
}
\label{fig:img}
\end{center}
\end{figure*}

In the past several months, the Swift and HETE-II satellites have
discovered four GRBs whose short duration ($t < 2$s) and spectral
hardness place them within the short-hard GRB
classification \citep{gehrels05,gcn3570,gcn3665,gcn3793}.  Furthermore,
each of these GRBs has been localized by its afterglow X-ray emission
to within a circle of radius 10\arcsec\ on the
sky \citep{bpp+05,ffc+05,bgk+05,mbk+05}. Although previous missions
reported hundreds of short-hard GRBs, none of these were promptly
localized to less than a few arcminutes and so a counterpart
association at other wavelengths proved elusive \citep{hbc+02,ngpf05}. 
The discovery of GRB\,050509b and a fading X-ray 
afterglow \citep{gehrels05} led to the first redshift and host 
galaxy association \citep{bpp+05} for a short-hard GRB, providing
unique insights for the long-standing mystery over the distance scale and energetics for at least some members of this class. The four events now localized offer an opportunity to (1) study the population of host galaxies and
large-scale environments, (2) examine the burst energetics, and 
(3) further constrain the nature of the progenitors.

In this Letter, we present imaging and discovery spectra of 
the galaxies hosting the short-hard GRBs 050509b, 050724, and 050813.
Based on these data, we report on their redshift, luminosity,
spectral-type, age, metallicity, and star-formation rates.
We also present a high-resolution spectrum 
($R\equiv\lambda/\Delta\lambda\sim 6000$) of the host of the
short-hard GRB~050709 and discuss its spectral 
properties and star formation characteristics.
We draw comparisons to the larger dataset of galaxies hosting long-soft
GRBs and discuss the implications for the progenitor origin of
short-hard GRBs.

\section{Observations and Analysis}

Optical images of the fields surrounding GRB\,050724 and GRB\,050813 
were obtained using the Gemini Multi-Object Spectrograph (GMOS) on 
the Gemini North Telescope with the $i'$ filter.  Optical images of
the fields surrounding GRB\,050509b and GRB\,050709 were obtained 
using the Echellette Spectrometer and Imager 
\citep{sheinis00} on the Keck\,II Telescope with the $R$ filter.  All
imaging data were taken under photometric conditions and were 
processed using standard IRAF tasks.  Photometric solutions were
derived from either a standard field taken during the night 
or from comparisons with USNO stars found in the science field.
Figure~\ref{fig:img} presents regions surrounding the localized
position of each short-hard GRB.
The processed images were registered to an absolute world 
coordinate system with typical 
1 $\sigma$ rms uncertainties of 150 milliarcsecond in each coordinate. 
We list the absolute positions of host galaxies 
for 050509b, 050709, and 050724 in Table~\ref{tab:summ}. 
We also list the magnitudes of the galaxies determined from our
imaging (converted to $R$-band magnitude for consistency)
with the exception of GRB~050509b where we report the
more accurate Sloan Digital Sky Survey $r'$ photometry \citep{sdssdr3}.

The ellipses in each panel represent the astrometric position of the 
most accurate X-ray afterglow position reported (68\% confidence 
interval for GRB\,050509b \citep{bpp+05}; 68\% confidence interval for 
GRB\,050709 \citep{ffc+05}; 68\% confidence interval for 
GRB\,050724 \citep{bgk+05}; 
and reflect the uncertainty in the astrometric tie 
between the X-ray and optical frame. 
A discussion of
the differences between the various determinations of the 
GRB~050509b XRT afterglow position has 
been discussed elsewhere \citep{bpp+05,pedersen06}.
The 90\% containment radius previously 
reported for GRB\,050813 \citep{mbk+05} is shown as a large circle. 
We have re-analyzed the X-ray data using an optimized technique for 
faint transient \citep{bpp+05} and localized GRB~050813 to
$\alpha$(J2000) = 16\fh 07\fm 56\fs .953 $\pm$ 0.20 sec, 
$\delta$(J2000) = +11\fd 14\arcmin 56\arcsec 60 $\pm$ 1.45 arcsec. 
The smaller ellipse shows this 68\% containment radius. 
This localization makes the host identification of 
$B$ or even the fainter $B*$ more likely over galaxy $C$.
We note that galaxies 
X, 
B, 
and C show  
consistent, red colors that suggest a cluster membership \citep{gcn3798}. 
The brightest objects at the edge of the large error 
circle (16\fh 07\fm 57\fs .393 +11\fd 14\arcmin 42\arcsec .79 and 16\fh 07\fm 56\fs .850 +11\fd 15\arcmin 01\arcsec .12) are 
likely foreground Galactic stars. 

\begin{figure*}
\begin{center}
\includegraphics[width=6.5in]{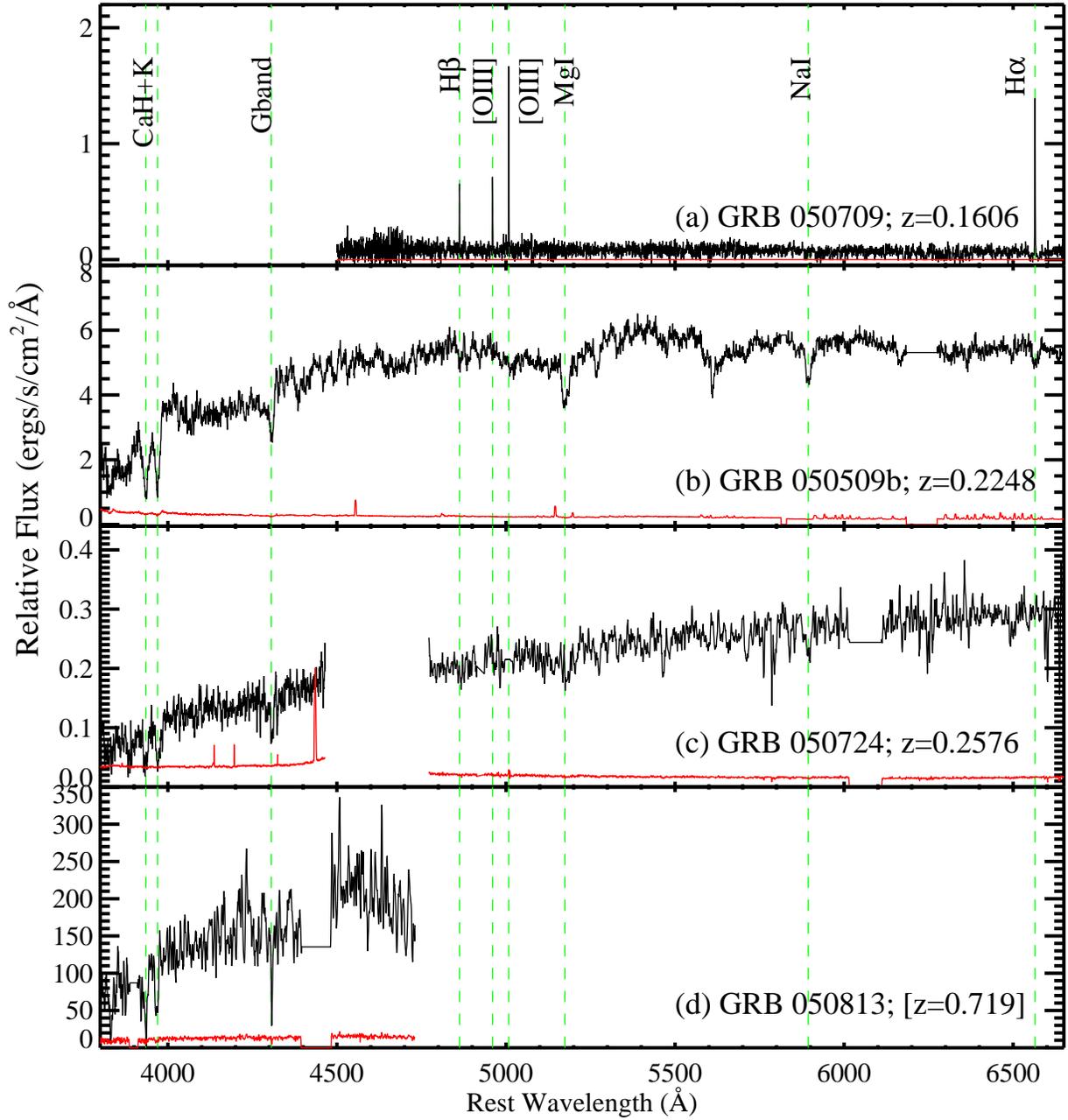}
\caption{
Optical spectroscopy for the host galaxies of short-hard GRBs.
With the exception of GRB~050724, these data are the discovery
spectra which established the redshift of the GRB event and also
the physical properties of the galaxy host and/or environment.
For GRB~050813, we show the spectrum for galaxy B.
}
\label{fig:spec}
\end{center}
\end{figure*}

Spectroscopy of the host galaxies were acquired on a number 
of facilities with several spectrometers (Figure~\ref{fig:spec}).  
Optical spectra of the host for GRB~050709 were obtained using
the Echellette Spectrometer and Imager 
on Keck~II with a $1''$~slit in echellette mode. 
Optical spectra of the host for GRB~050509b
were obtained using the DEIMOS spectrometer 
on Keck~II with a $0.7''$ longslit and the 600line/mm grating.
Optical spectra of the host for GRB~050724 were obtained using
the LRIS spectrometer on 
the Keck~I telescope with the 600/4000 grism
through a $1''$ longslit for $\lambda < 4500$\AA\ and using
the GMOS spectrometer on the Gemini-North telescope with a $0.75''$ slit 
(following astrometry based on a Magellan guide-camera image) 
and the R400 grating centered at 690nm providing spectra with
$\lambda > 4700$\AA.  Optical spectra of galaxies B,C, and X in
the field surrounding GRB~050813 were obtained using
the GMOS spectrometer with the same instrumental setup that was applied
for the host GRB~050724 except centered at 640nm.
The data were fluxed using spectrophotometric standards taken with the
same instrumental setups. The absolute flux is an underestimate,
however, due to slit losses and reddening by the Milky Way. 

The redshifts of the galaxies were measured 
through fits to the spectral features indicated in the figure. 
Based on these redshifts, we have measured or constrained the star
formation rate (SFR) for these galaxies by measuring the luminosity
of H$\alpha$ and/or [OII].  In the cases of GRB~050509b, GRB~050724,
and GRB~050813, we do not detect any significant emission and we
employ the SFR calibration of \citep{ken98} to place conservative
upper limits on the current star formation rate (Table~\ref{tab:summ}).
In the host galaxies of GRB~050509b and GRB~050724, the absence of 
strong H$\beta$ absorption also indicates that there has been no
significant SFR over the past $\approx 1$~Gyr.

In the case of GRB~050709, there are strong emission lines observed
indicating significant ongoing star formation \citep{gcn3605,ffc+05}.  
We have estimated
the SFR by first comparing the relative H$\alpha$ and H$\beta$ fluxes
to measure the reddening along the sightline 
to the galaxy under the assumption of Case~B recombination.  
We infer a reddening $E(B-V) > 0.4$ and de-extinct the
H$\alpha$ emission to derive a luminosity 
$L_{\rm H\alpha} > 4 \times 10^{40} {\rm ergs/s}$.  We report this
as a lower limit because 
(i) we have applied only a conservative correction to the H$\alpha$
flux due to the B-band absorption; and 
(ii) the slit does not encompass the entire galaxy. 
Using the \cite{ken98} empirical relation, we derive a lower limit
to the SFR ${\rm SFR} > 0.3 \msol {\rm yr^{-1}}$ in reasonable
agreement with other estimates \citep{ffc+05,covino05}.
We do not, however, confirm the absorption features at H$\alpha$
and H$\beta$ reported by \cite{covino05} and have no sensitive
age constraint for this galaxy.

We present only the spectrum for galaxy~B associated 
with GRB~050813 (Figure~\ref{fig:img}).  Our spectrum of galaxy~C
shows a 4000\AA\ break consistent with $z=0.73$ and no significant
emission lines, galaxy~X shows absorption features indicating
$z=0.722$ \citep[see also][]{gcn3801}, and we have no redshift
constraint for galaxy $B*$ ($i = 24.2 \pm 0.1$).
The small projected
distance between these sources ($\approx\,40-100\,h_{70}^{-1}$ kpc)
and large velocity difference ($\Delta v = 690-3000$ \kms) strongly
support the cluster nature of the progenitor environment for
GRB050813 \citep{gcn3798}. 
We have also obtained spectra of two bright galaxies near GRB\,050724 
at positions 16\fh 24\fm 46\fs .739 $-27$\fd 32\arcmin 28\arcsec .90 
and 16\fh 24\fm 43\fs .344 $-$27\fd 32\arcmin 07\arcsec .21. 
The latter galaxy has a redshift $z=0.316$ based in H$\beta$ and
H$\alpha$ emission.  The former galaxy shows no absorption or
emission features consistent with $z=0.2576$ and its spectrum
suggests a redshift $z>0.4$.
We therefore have found no evidence that GRB\,050724 is located within
a galaxy cluster.

We have constrained the age and metallicity of each early-type galaxy by 
performing least-squares fits to the galaxy spectrum with
a suite of idealized templates \citep{bruzual03}.  For the templates, 
we adopted a Salpeter initial mass function and considered a range 
of metallicity from 0.2 solar to solar, as well as different star 
formation histories, ranging from a single burst model to an 
exponentially declining SFR with an e-folding time of 0.3 Gyr. 
None of the galaxy spectra are well matched by the burst model.
The analysis places tight constraints ($<30\%$ uncertainty) on the metallicity 
of each galaxy, but the inferred age is more uncertain 
especially because it is sensitive to the reddening assumed.  
With the exception of GRB~050724 where we correct for Galactic extinction
\citep[$E(B-V)=0.61$][]{schlegel98}, we assume no 
reddening.  To be conservative, we report minimum ages for the galaxies
(Table~\ref{tab:summ}).  The values compare favorably with other estimates
for GRB~050724 \citep{gorosabel05}, but our results contradict the
young age for GRB~050509b reported by \cite{castro05}.
Note that the galaxy spectra associated with GRB~050813 do not 
have sufficient spectral coverage to provide a meaningful age constraint.
Finally, we estimate the metallicity for the galaxy associated with
GRB~050709 from the measured [\ion{N}{2}]~$\lambda 6583$ flux
relative to H$\alpha$ using the empirical calibration given
by \cite{pp04}.  We measure log([\ion{N}{2}]/H$\alpha$)~$= -1.3 \pm 0.1$
which implies log(O/H)~$= 8.16 \pm 0.06$, comparable to the
low metallicities observed in long-soft GRB host galaxies
\citep[e.g.][]{pro04}.

\section{Discussion}

Based on positions of the afterglows, two of four bursts 
(050509b and 050813) are very likely
associated with clusters of galaxies \citep{bpp+05,gcn3798}. Because
only $\approx 10$\% of the mass of the Universe is 
contained within massive clusters, this suggests that either galaxies in clusters 
preferentially produce progenitors of short-hard GRBs or that 
short-hard bursts are preferentially more likely to be 
localized in cluster environments \citep{bpp+05}. We have examined the 
Swift X-ray Telescope data of the fields of the other two GRBs (050709 
and 050724) and found no conclusive evidence for diffuse hot gas 
associated with massive clusters. Furthermore, a spectroscopic study 
of two bright galaxies near the X-ray afterglow position of GRB\,050724 
show them at different redshifts, disfavoring a cluster 
origin for that burst.  The cluster environments of at least two 
short-hard GRBs contrast strikingly with the observation that no 
well-localized long-soft GRB has yet been associated with a 
cluster \citep{bml+04}. Therefore, 
more sensitive observations of the fields 
of both historical and new well-localized short-hard GRBs may
be expected to
show a significant preponderance to correlate with galaxy clusters.

We now turn to the putative galaxy hosts of short-hard GRBs. In three of 
four cases, the GRB has been plausibly associated with a galaxy to better 
than a 99\% confidence level (Figure\ref{fig:img}). In the fourth case
(050813), there are two 
galaxies located in the error circle with comparable magnitude and one 
may associate the event with either of these. 
Three of the bursts are associated with galaxies exhibiting characteristic 
early-type spectra (Figure~\ref{fig:spec}). 
The absence of observable H$\alpha$ and [O\,II] 
emission constrains the unobscured star formation rates (SFR) in 
these galaxies to ${\rm SFR} <0.2 \msol {\rm yr^{-1}}$, 
and the lack of Balmer absorption lines implies 
that the last significant star forming event occurred $>1$ billion years 
ago.  The host galaxy of GRB\, 050709 exhibits strong emission lines 
that indicate on-going star formation with a conservative lower limit 
of ${\rm SFR} >0.5 \msol {\rm yr^{-1}}$. These observations 
indicate that these short-hard GRBs occurred during 
the past $\sim 7$ billion years of the 
Universe $(z<1)$ in galaxies with diverse physical characteristics. 

In contrast to what is found for short-hard GRBs, all of the 
confirmed long-soft GRB host galaxies are actively forming stars with 
integrated, unobscured ${\rm SFRs} \approx 1 - 10 \msol {\rm yr^{-1}}$ 
\citep{chg04}. 
These host galaxies have small stellar masses and bluer colors than 
present-day spiral
galaxies \citep[suggesting a low metallicity][]{ldm+03}.  
The ages implied for the long-soft GRBs are estimated to
be $<0.2$~Gyr \citep{chg04} which are significantly younger than the
minimum ages derived for the early-type galaxies in our sample
(Table~\ref{tab:summ}).
We conclude that the host galaxies of short-hard GRBs, and by extension 
the progenitors,  are not drawn from the same parent population of 
long-soft GRBs. And although long-soft GRBs are observed to 
significantly higher redshift than the current short-hard GRB sample,
one reaches the same conclusions when restricting to low-$z$,
long-soft GRB hosts \citep{sof+05}.

The identification of three galaxies without current star formation 
argues that the accepted progenitor model of long-soft GRBs (the collapse 
of a massive star; \citealt{woo93}) is not tenable as a source for the 
short-hard GRBs. Instead,  the observations lend support to theories 
in which the progenitors of short-hard GRBs are merging compact binaries 
(neutron stars or black holes \citep{pac86,elps89}). This inference 
is supported through several channels. First, the redshift distribution 
of these short-hard bursts is inconsistent with a bursting rate that 
traces the star-formation rate in the universe, unlike long-soft GRBs, 
which do follow it. If we introduce a $\sim 1$\,Gyr time delay 
from starburst to explosion, as expected from compact object mergers, 
the observed redshift distribution of these GRBs (i.e.\ assuming they
are representative of short-hard GRBs in general)
is consistent with the 
star-formation rate \citep{gp05}.  Second, the lack of an associated 
supernova for all four short-hard GRBs is strong evidence against a 
core--collapse origin \citep{bpp+05,hsg+05}.  
Third, our measured offsets (Figure~\ref{fig:img}) 
of the short-hard GRBs from 
their putative hosts are compatible with predicted site of merging compact
remnant progenitors \citep{fwh99,bsp99}. 
This includes the small offset 
of GRB\,050724 (2.36 $\pm$ 0.90\,kpc) which
is near the median predicted merger offset for such galaxies \citep{bsp99}.

The identification of the host galaxies and redshifts fixes
the isotropic-equivalent burst energies. Table~\ref{tab:summ} 
shows the inferred 
isotropic energy release in prompt $\gamma$--ray emission, along with 
its duration in the source  rest--frames. These events suggest that 
short-hard GRBs
are less energetic, typically by more than one order of magnitude, than 
their long counterparts, which typically release a total $\gamma$-ray 
energy of $5 \times 10^{50}$~erg when collimation is taken into 
account.  The total isotropic-equivalent energy in 
$\gamma$-rays, $E_{\gamma,{\rm iso}}$ appears to
correlate with the burst duration, such that longer events are also
more powerful \citep{bpc+05}. 
We find that $E_{\gamma, {\rm iso}} \propto T_{90}^{\psi}$ and
$\psi\approx 3/2$ to 2.  The total energies, durations, and the general
behavior of the correlation between them are in rough agreement with
the numerical modeling of GRB central engines arising from compact
object mergers \citep{lrrg05,oj05,rrd03}. 


The association of short-hard GRBs with both star-forming galaxies 
and with ellipticals dominated by old stellar populations is analogous to 
type Ia SNe.  It indicates a class of progenitors with a wide 
distribution of delay times between formation and explosion, with 
a tail probably extending to
many Gyr.  Similarly, just as core-collapse supernovae are discovered 
almost exclusively in late-time star-forming galaxies, so too are 
long-soft GRBs.
The detailed physics of the progenitors of supernovae is inferred
through the time evolution of metals and ionic species revealed by
spectroscopic observations. However, the progenitors of GRBs are
essentially masked by afterglow emission, largely featureless
synchrotron light, which reveals little more than the basic energetics
and micro-physical parameters of relativistic shocks.  
As new redshifts, offsets and host galaxies of short-hard GRBs 
are gathered, the theories of the  progenitors will undoubtably 
be honed.  Still, owing to the largely featureless light of 
afterglow radiation, unless short-hard bursts are eventually 
found to be accompanied by tell-tale emission features like 
the supernovae of long-duration GRBs, the only definitive 
understanding of the progenitors will come with the observations 
of concurrent gravitational radiation or neutrino signals arising 
from the dense, opaque central engine.

\acknowledgements
We thank S. Sigurdsson and D. Kocevski for useful discussions.
Some of these observations were made with the W.M. Keck Telescope.
The Keck Observatory is a joint facility of the University
of California, the California Institute of Technology, and NASA.
We are grateful to the staff of Gemini for their assistance in 
acquiring this data. J.X.P., J.S.B., and H.-W.C. are
partially supported by NASA/Swift grant NNG05GF55G.
ERR and D.Pooley were sponsored by NASA through 
Chandra Postdoctoral Fellowship awards PF3-40028 and PF4-50035.
Work at LLNL is performed under the auspices of the U.S. Department of
Energy and Lawrence Livermore National Laboratory under contract No.
W-7405-Eng-48. Based in part on observations obtained at 
the Gemini Observatory, which is operated by the 
Association of Universities for Research in Astronomy,
Inc., under a cooperative agreement with the NSF on behalf of the
Gemini partnership: the National Science Foundation (United States),
the Particle Physics and Astronomy Research Council (United Kingdom),
the National Research Council (Canada), CONICYT (Chile), the Australian
Research Council (Australia), CNPq (Brazil) and CONICET (Argentina).

\bibliographystyle{apj}

\begin{table*}
\begin{center}
\caption{{\sc Physical Characteristics of 
Short-Hard GRBs and their Putative Host Galaxies\label{tab:summ}}}
\begin{tabular}{lcccccrcl}
\tableline
\tableline
Property & 050509b & 050709 & 050724
& 050813(B) & 050813(C) & 050813(X) \\
\tableline
T$_{90}/[1+z]$ (sec)$^a$ & 0.032 & 0.060 & 0.20 & 0.35 & 0.35 & 0.35 \\
$E_{\gamma,{\rm iso}}$(erg)$^{b}$ &  
$2.75 \times 10^{48}$ & $2.29 \times 10^{49}$ & $1.0 \times 10^{50}$ 
& $1.7 \times 10^{50}$ & $1.7 \times 10^{50}$ & $1.7 \times 10^{50}$ \\
$\alpha$(J2000) & 12:36:12.878 &  23:01:26.849 & 16:24:44.381 & 
16:07:57.200 & 16:07:57.008 & 16:07:57.509 \\
$\delta$(J2000) &  +28:58:58.95 & $-$38:58:39.39 & $-$27:32:26.97 & 
+11:14:53.09 &  +11:14:47.37  & +11:15:02.13 \\
$z_{gal}$ & $0.2248 \pm 0.0002$ & $0.1606 \pm 0.0001$ & $0.2576 \pm 0.0004$ & 
 $0.719 \pm 0.001$ & $0.73 \pm 0.01$ & $0.722 \pm 0.001$   \\
$r^c$ (kpc) & $39 \pm 13$ & $3.5 \pm 1.3$ & $2.4 \pm 0.9$ \\
$R^d$ (mag) & $16.8 \pm 0.05$ & $21.1 \pm 0.2$ &  $19.8 \pm 0.3$ & $23.43 \pm 0.07$ 
& $22.57 \pm 0.07$ & $22.75 \pm 0.07$ \\
$L_B^e$ ($10^9 L_\odot$) & 100 & 1.5& 8.5 & 8 & 18 & 15 \\
SFR$^f$ ($M_\odot {\rm yr^{-1}}$) & $<0.1$ & $>0.3$ & $<0.05$ & $<0.1$ & $<0.2$ & $<0.1$ \\
Metallicity$^g$ ($Z/Z_\odot$) & 1 & 0.25 & 0.2 & 1 & ... & 1 \\
Min. Age (Gyr) & 3 & ... & 8 & ... & ... & ... \\
Spectral Type &  Elliptical &  Late-type dwarf &  Early-type &  Elliptical 
     &  Elliptical &  Elliptical \\
\tableline
\end{tabular}
\end{center}
\tablenotetext{a}{Source rest--frame duration, measured in $T_{90}$, the time 
when 90\% of the total fluence of the GRB is accumulated, beginning 
after 5\% of the fluence has been accumulated \citep{kmf+93}. Values
were reported by the Swift and HETE-2 Teams 
\citep{gcn3381,gcn3653,gcn3667,gcn3793}.}
\tablenotetext{b}{
Isotropic-equivalent energy $E_{\gamma,\rm iso}$, computed using 
the observed fluence and redshift under the assumption of a concordance 
cosmology with $\Omega_m = 0.29$, $\Omega_{\Lambda} = 0.71$ and Hubble's 
constant $H_0 = 70$ km s$^{-1}$ Mpc$^{-1}$. While these energies are 
systematically lower than for long-soft GRBs, we note that with the 
energy range covered by Swift (15--350~keV) and the spectral 
properties of the prompt emission, the derived values should be 
considered lower limits.}
\tablenotetext{c}{Projected offset of the X-ray afterglow positions
from the optical centroid of the respective host galaxies.
The quoted error is an approximation to the uncertainty of the 
most likely offset $r$, following appendix B of Bloom et al.\ (2002)
which is required because offsets are a positive-definite quantity 
and not strictly Gaussian. In general, $r \pm \sigma_r$ does not 
contain 68\% of the probability distribution function.}
\tablenotetext{d}{$R$-band magnitudes. We convert the Sloan Digital Sky
Survey $r$ magnitude for 050509b \citep{gcn3418}. 
For the galaxies associated
with GRB~050813 we have measured $i$-band magnitudes and
converted to $R$-band assuming $R-i = 0.99$\,mag, appropriate
for an elliptical galaxy at $z=0.7$.}
\tablenotetext{e}{The $R$-band magnitudes 
were converted to $B$-band luminosities 
by assuming standard colors for these spectral types, adopting
the redshift listed in column~1, and adopting the standard cosmology.
The luminosities have not been
corrected for Galactic extinction and are reported relative to the
Solar $B$-band luminosity.}
\tablenotetext{f}{Unextinguished star formation rate based on H$\alpha$ 
and/or [OII] luminosity.  Upper limits are 3$\sigma$.}
\tablenotetext{g}{Based on template fits to the galaxy spectra except
for GRB~050709 where we estimate (O/H) from the [\ion{N}{2}]/H$\alpha$ ratio
\citep{pp04}.  The uncertainty in these values is $<30\%$.}
\end{table*}

\end{document}